# THE STRING TENSION IN GAUGE THEORIES

E. MARINARI[0], M. L. PACIELLO[1] and B. TAGLIENTI[1]

[0] *Dipartimento di Fisica ed Infn, Università di Cagliari*
*Via Ospedale 72, 09100 Cagliari, Italy*
[1] *Infn, Sezione di Roma, Università di Roma* La Sapienza
*P. A. Moro 2, 00185 Roma, Italy*

## 1. Introduction

Lattice gauge theories[1,2] have allowed a deep understanding of the continuum behavior of non-abelian gauge theories. It was not thinkable, before the introduction of such a powerful non-perturbative tool, to have such a good control of properties like quark confinement and of the gluonic and fermionic mass spectrum.

The string tension, as defined in the quenched theory, which we will mainly discuss in the following, is maybe the main non-perturbative quantity. We use it to learn that the theory is confined, and to gather non-perturbative information about the behavior of quark pairs.

Monte Carlo simulations have been described in a very complete manner at various stages of their developments, both in reprint collections[3], in books dedicated to the subject[4,5,6,7] and in reviews[8,9,10]. The Lattice conferences proceedings provide an invaluable source of information. We will not quote them one by one, but address the reader to the yearly reviews on the string tension and, more in general, to the detailed contributions contained in the proceedings.

Also thanks to the existence of such reviews here we will not have to condense large series of numbers, but we will just try to reconstruct a general frame, and to stress a few main ideas, and accomplished and yet to come developments. We will try here to give the basics that a beginner would need to approach the field, to stress a few ideas that look important to us, and to indicate the direction of the most promising developments. Nothing more.

In section (2) we will introduce Wilson loops, and their use to measure the string tension. We will discuss there the first string tension measurements. In section (3) we introduce Polyakov loops. In section (4) we discuss the continuum limit, and how one gets sure that she is measuring a quantity that does not depend anymore on the specific (lattice) discretization. In section (5) we discuss the measurements of the quark anti-quark potential. Since this is a crucial physical point we will go here in some detail.

Maybe the main point of the review is in section (6), where we discuss a variety of methods to measure the string tension. This has been one of the crucial developments which have allowed to use the computers to extract non trivial physical predictions. In section (7) we discuss about universality, and, again, about which is the best to the continuum limit. In the section (8) we discuss, in some sense, about finite size effects. It is interesting that the picture of a confinement given by a thin flux confined to a tube can lead to quantitative predictions of the size of the non-dominant terms. In section (9) we draw our conclusions.

## 2. Wilson loops

In this section we will discuss the use of Wilson loops to measure the string tension in non-abelian gauge theories. This is the simplest framework in which non-perturbative quantities can be computed (at least numerically) with good precision. Wilson ideas[2] find here a practical implementation, and become operative tools. In the following we will discuss mainly about the quenched theory (where fermions do not appear in the functional measure). Here fermions are external sources, which we use as a probe to test the intrinsic confining properties of the gauge field configurations which contribute to the path integral.

One of the main goals of a non-perturbative formulation of Gauge Theories is trying to understand the phenomenon of quark confinement. In a non-abelian gauge theory color sources are confined in color-singlets, which cannot be separated. At a qualitative level such confinement is understood by assuming that the color electric fluxes emanating from the quarks are squeezed into a string-like configuration. Such a gauge field configuration has a constant energy density per unit length, i.e. $E = \sigma r$. This energy density $\sigma$ is called the *string tension,* and it is the fundamental physical quantity for a pure gauge field configuration (in the full theory the string can break by creating a quark-anti-quark pair, and we have to use another criterion to check confinement). Verifying such a scenario has been one of the major tasks of numerical simulations of lattice gauge theories.

The fundamental quantity one defines when formulating Lattice Gauge Theories is the Wilson loop $\langle W_\gamma \rangle$, which will also be one of the basis of our string tension measurements. Let us consider a rectangular closed path $\gamma$ in ($4d$ in our case) Euclidean space-time, extending for a length $x$ along one of the 3 spatial directions and for a length $t$ along the time axis. For the $SU(N)$ gauge theory on defines by $\langle W_\gamma \rangle$ the expectation value of $\frac{1}{N} Tr \, U_\gamma$ where $U_\gamma$ is the transport operator along the path $\gamma$ (in the fundamental representation of the gauge group). On a 4 dimensional hypercubical lattice $U_\gamma$ is simply given by the product $U_{i_1 i_2} U_{i_2 i_3} .... U_{i_N i_1}$ of the (oriented) dynamical variables associated with the links joining the consecutive lattice sites one finds along the paths.

In the limit of large separation $\langle W_\gamma \rangle$ describes the change in the vacuum to vacuum transition amplitude induced by the presence of an external current. The current corresponds to the creation at some initial time of a pair of static sources at separation $x$, the propagation of the sources for a time interval $t$ and the final annihilation of the sources. Wilson loops are a probe for quarks. From their large separation behavior one learns about the behavior of hypothetical (infinitely heavy) quark when put in an typical equilibrium gauge field configuration. As we will see a confining scenario implies an *area law* decay of the Wilson loops expectation values.

If we expand over a complete set of eigenstates of the Hamiltonian (in presence of an external source) $\langle W_\gamma \rangle$ can be expressed as:

$$\langle W_\gamma \rangle = \sum_n \, C_n \, e^{-V_n t} \,, \qquad (2.1)$$

where the $C_n$ are the matrix elements for the creation of the sources from the vacuum and the exponential factors are real because of the Wick rotation to imaginary time (see for example [9] and [11]).

For large $t$ the state with the lowest energy eigenvalues $E_0$ dominates the statistical sum. From the large $t$ and $x$ decay of $\langle W_\gamma \rangle$ one can infer the behavior of the energy density.[*]

Typically one finds that at large $x$ and $t$

$$\langle W_\gamma \rangle \approx e^{-\sigma x t - m(x+t) + \cdots} \,, \qquad (2.2)$$

with an area and a perimeter term. The perimeter contribution $e^{-m(x+t)}$ is suggested from the form of the self-energy term in the perturbative computation of the heavy quark potential (see for example [9]). Now one measures numerically the expectation value of $W_\gamma$, by the standard sampling techniques. If asymptotically one finds that the Wilson loop decays according to an area law, i.e.

$$\langle W_\gamma \rangle \approx e^{-\sigma x t} \,, \qquad (2.3)$$

---

[*]It should be always clear that the definition of the string tension obtained from the asymptotic behavior of the quark-quark potential is in principle different from the one obtained by using directly the energy of the electric flux tube, and that the two have to be equal only asymptotically. For example the size of the string fluctuations can be different in the two cases. See for example ref. [12] for an interesting discussion of the subject, and references therein.

than the theory is confined. The effective potential is asymptotically linearly increasing, and one estimates the value of $\sigma$ by such a decay rate. On the contrary a theory with free particles gives raise to a perimeter decay law. That is for example what happens for electrons in the $U(1)$ gauge theory (in the continuum limit).

Let our rectangular loop

$$\langle W_\gamma \rangle \equiv W(I, J) \tag{2.4}$$

(on a hypercubical lattice in four Euclidean space-time dimensions) have a $I \times J$ size ($x = Ia$, $t = Ja$, where $a$ is the value of the lattice spacing). As $x \to \infty$ $W(I, J)$ can be expressed in terms of the potential $V(x) \approx \sigma x = \sigma Ia$ as

$$W(I, J) \approx e^{-\sigma a^2 IJ} \ . \tag{2.5}$$

A lattice calculation of $W(I, J)$ will not produce $\sigma$ directly but a dimensionless (since we always measure pure numbers) function $K$, depending on the lattice coupling constant $g$ through the lattice spacing $a$, related to $\sigma$ by

$$K(g) = \sigma a^2 \ . \tag{2.6}$$

In the following we will also use the variable $\beta = 2N/g^2$ for the $SU(N)$ gauge groups. Section (4) is devoted to study this dependence.

To measure $\sigma$ Creutz introduced the ratio[14]

$$\frac{W(I, J)W(I-1, J-1)}{W(I, J-1)W(I-1, J)} \ . \tag{2.7}$$

This ratio has the advantage of making more clear the area dependence of the Wilson loops. If $W$ behaves as we have supposed in (2.2), then the Creutz ratio has the value

$$\exp -\sigma a^2 \ . \tag{2.8}$$

Obviously for small loops short distance corrections will spoil the behavior (2.8). One will have to measure loops that are *large enough* (in some sense we will discuss better in the following) and check a posteriori, self-consistently, the presence of a confining behavior and of an area law decay. Technically one takes the logarithm of the ratio's (2.7), and defines

$$\chi(I, J) = -\ln \frac{W(I, J)W(I-1, J-1)}{W(I, J-1)W(I-1, J)} \ . \tag{2.9}$$

When measuring these quantities difficulties arise from the finite extent of the lattices. Moreover when $W(I, J)$ becomes too small the error due to statistical fluctuations makes the measurements meaningless: on a finite lattice, for finite $I$ and $J$, one always estimate a value of $\chi$ larger then the true value of $\sigma a^2$. To extract the string tension one has to use the envelope of the curves $\chi(I, J)$. Let us stress that in the $\chi$ ratios we have canceled only the dominant perimeter behavior of (2.2). Contributions like for example a term $e^{\frac{J}{T}}$ (expected from the fluctuations of the *surface* of the path[9]) do survive, and only cancel in the asymptotic limit. We will give a more accurate analysis of the quark-antiquark potential behavior in section (5). For a pedagogical review, containing explicit examples about the quark potential and the asymptotic freedom in continuum quantum cromodynamics, see[9].

First numerical results for the $SU(2)$ and $SU(3)$ gauge theory were presented in [14,15,16,17,18,19]. Computations for different lattice sizes have been performed and the dependence of the results over the Euclidean time extension of the lattice has been analyzed.

From such first string tension measurements it became clear the crucial role of the control of finite volume and finite Euclidean time effects, of the contamination from excited states and the distortion of the spectrum due to the finiteness of the lattice. In particular we will see that the difficult problem of extracting the rate of an exponential behavior out of numbers of order one has been overcome not only with more powerful computers, but (and maybe this has been the crucial step) also by developing effective numerical algorithms.

### 3. Polyakov loops

Polyakov lines, which wrap the periodic lattice in one direction, have been first used in the context of Lattice Gauge Theories, in ref. [20] and [21], to monitor a finite temperature phase transition. Polyakov lines turn out to be, as we will discuss in the following, very useful tools to get reliable measurements of the string tension.

After the first string tension measurements, which we have discussed in the previous section, larger scale measurements of the Wilson loops, at lower values of the inverse coupling constant $\beta$, suggested that the asymptotic string tension could assume a value substantially smaller than previously determined.

It was remarked in[17] that a potentially effective way of measuring the string tension was based on studying the large separation behavior of correlation functions of two Polyakov loops (see fig. 1).

Denoting by $P(x, y, z)$ the trace of the product of the link gauge fields in the $t$ direction, closed thanks to the periodic boundary conditions (the so called *Polyakov loops*), let us consider in a $4d$ lattice, with $N_t$ sites in the $t$ direction. the correlation functions of two Polyakov loops wrapping the lattice in the $t$ direction, separated in the $z$ direction by $z$ sites. In the scaling limit, for $z$ large enough and $N_t \to \infty$, one recovers the quark-anti-

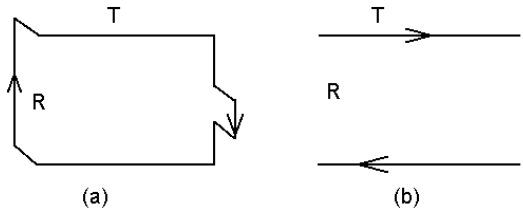

Figure 1: (a) Generic correlation of Wilson loops of length $R$ separated by the time distance $T$; (b) Correlation of Polyakov loops encircling the periodic time direction.

quark potential by

$$\langle P(0,0,0)P(0,0,z)\rangle_c \simeq_{N_t \to \infty} e^{-N_t \ V(z)} \ . \tag{3.1}$$

If in the large $z$ region $V(z)$ increases linearly with the distance, i.e. $V(z) \approx Kz$, then

$$\langle P(0,0,0)P(0,0,z)\rangle_c \simeq_{N_t \to \infty} e^{-N_t \ Kz} \ . \tag{3.2}$$

In other words $(N_t \ K)$ is the *mass* of the correlation function of two Polyakov loops.

The coefficient $(Kz)$ which governs the decay of the correlation function can be seen as the energy $E_0(z)$ of a string of length $z$ which propagates for an (Euclidean) time $N_t$. As for a glueball state (see later), the mass at rest of the system can be extracted by summing over spatial planes to project out the zero momentum component of the state. Since the straight Polyakov loop is already symmetric in the $z$ direction, only an average over the remaining two directions is needed. The quantities considered when measuring the $SU(3)$ string tension[22,23,24,25,26,27,28] are correlations of zero momentum sums of Polyakov loops:

$$C(z) \equiv \langle \sum_{x,y}[P(x,y,0)] \sum_{w,v}[P(w,v,z)]\rangle_c \to_{N_t \to \infty, z \to \infty} e^{-N_t \ Kz} \ . \tag{3.3}$$

For a lattice of sufficiently small size $N_z$ in the $z$ direction there will be a deconfining phase transition (a first order transition for $SU(3)$[29,30,31,32] and a second order transition for $SU(2)$[20,21,31,33,32]). In this gluon plasma phase the correlation between two Polyakov loops is given by an inverse power law rather than by an exponential because of the zero mass deconfined gluon modes. We use fig. 2 to illustrate the situation.

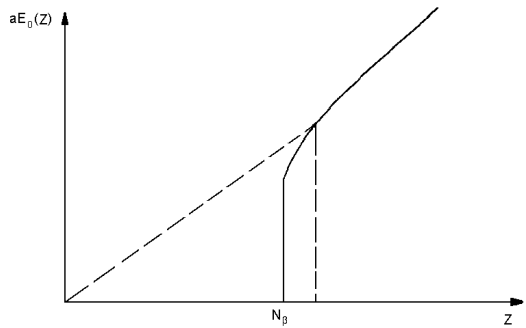

Figure 2: The expected form of the energy $E_0(z)$ as function of $z$: $E_0(z) = 0$ for $z \leq N_\beta$ where $N_\beta$ is the finite temperature transition in a $N_\beta \times \infty^3$ lattice. The extracted value of $E_0(z)/z = K_{eff}$ leads to $K_{eff} \leq K$ where $K$ is the asymptotic slope of $E_0(z)$

The string tension extracted from a finite lattice will be smaller than the true string tension given by the asymptotic slope of $E_0(z)$ in $z$. The presence of such finite temperature effect makes clear the fact that $E_0(z)$ cannot be linear with $z$. Considering contributions of different nature is needed for achieving a complete and satisfactory description.

The general requirement needed for obtaining a fair measurement of the string tension is to use a lattice large enough so that the contribution from the fake quark loops wrapping around the lattice are sufficiently small. This effect is connected to the finite temperature phase transition which is triggered when such a contribution becomes significant. In fig.3, from ref.[76], we show the minimum lattice size, as a function of $\beta$, needed to avoid finite size effects for $SU(3)$.

A complete review of the measurements of SU(2) and SU(3) string tension until 1987 can be found in ref.[76]; more recent data can be found for SU(3) in ref.[60,58,69], for SU(2) in ref.[100,58,63,59] and in the references discussed in more detail in the following sections.

An interesting comparison between the string tension measured from Wilson loops and Polyakov loops can be found in references[34,35]. In particular ref.[35] shows that using the results of a simplified string model for the confined pure gauge field theory it is possible to explain some of the systematic effects in the measurement of the lattice QCD string tension. In fact, the finite lattice used in computer simulations causes the string tension to be overestimated when using Wilson loops and to be underestimated when using Polyakov loops.

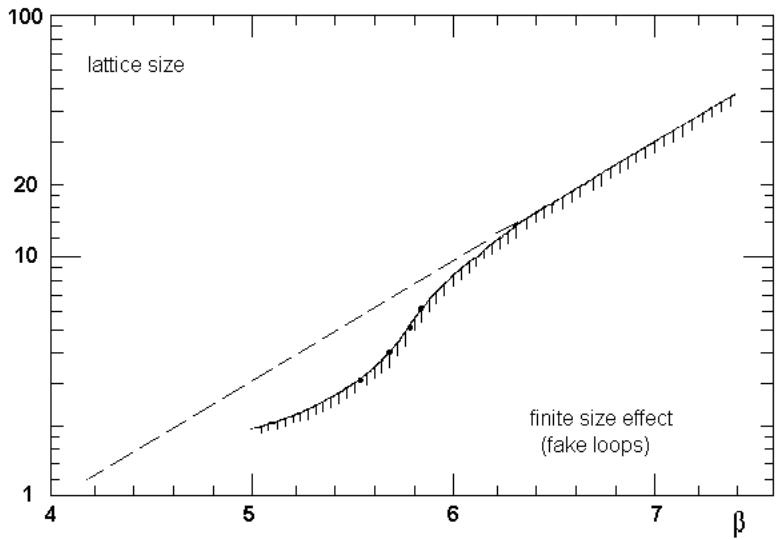

Figure 3: Minimum lattice size estimated from the critical coupling of the finite temperature phase transition as a function of $\beta$ from ref.[76].

Let us finally remark that Polyakov loops method can be generalized to the case of multi-quark systems to extract the static energy of the system containing quarks and antiquarks[36,26,37].

### 4. Continuum limit, scaling regime and $\beta$-function

We want now to discuss how from a lattice theory one can recover a continuum theory. We are interested in continuum QCD, the non-abelian theory which we expect to describe strong interactions, and the lattice theory is for us a very useful tool, since it provides a gauge invariant non-perturbative renormalization scheme for the continuum gauge theory. The lattice theory is gauge invariant for all values of the lattice spacing, this is the big bonus we get thanks to the Wilson formulation. The fact that we are dealing with an asymptotically free theory (i.e. that quarks behave as free at very short distances)[38,39] implies that in the lattice model we need to send $g \to 0$ to obtain the continuum limit (of zero lattice spacing, $a \to 0$). The point that explains this fact can be visualized by thinking at a finite *physical* interaction range. Continuum limit means that the lattice grid is irrelevant, i.e. in a constant physical distance (given by the interaction range) we have more and more grid points. But that happens when the interaction becomes weaker, i.e. in our theory in the short distance limit.

In some sense QCD is a happy and easy case, since the critical exponents can be computed in perturbation theory; around $g^* = 0$, the infrared fixed point[40] of the theory, perturbation theory and renormalization group are used to understand the critical behavior. Numerical simulations allow to extract the non-perturbative content of the theory. Then an analysis is performed on the *scaling region* of the theory, where the perturbative scaling regime is expected to hold.

To give a hint about the behavior of the mass gap of the theory, we use the lattice regularization noting that what we are saying would be valid in all schemes (but only the lattice scheme is *non-perturbative* and *gauge-invariant* and will allow the next, crucial step, numerical simulations).

One requires that a physical quantity with dimensions mass $M$ be a renormalization group invariant, i.e not to depend on the lattice spacing in the scaling regime which characterizes the approach to the continuum limit:

$$a\frac{dM}{da} = 0 \ , \tag{4.1}$$

where $a$ is the cutoff. Since $M$ has the dimensions of a mass, in the continuum limit, where the correlation length diverges in lattice units and there are no scales left (the lattice grid disappears in this limit) we have

$$M = a^{-1}\phi(g) \ , \tag{4.2}$$

where $\phi$ is a function of the coupling $g$. Substituting $M$ in (4.1) we see that $\phi'$ behaves as

$$\phi'(g) = -\frac{\phi(g)}{\beta(g)} \ . \tag{4.3}$$

Here $\beta(g)$ stands for the well known Callan-Symanzik $\beta$ function[41]

$$\beta(g) = -a\frac{\partial g}{\partial a} \ . \tag{4.4}$$

For non-abelian quantized gauge theories it is possible to compute in perturbation theory the functional form of $\phi(g)$ in eq. 4.3 in the vicinity of the fixed point $g^* = 0$. This leads, for a pure $SU(N)$ theory, to the result [38,39]:

$$a\frac{\partial g}{\partial a} = \beta_0 g^3 + \beta_1 g^5 + .... \tag{4.5}$$

with

$$\beta_0 \equiv \frac{11}{3}(\frac{N}{16\pi^2}), \ \beta_1 \equiv \frac{34}{3}(\frac{N}{16\pi^2})^2 \ . \tag{4.6}$$

Using eq. 4.5 the formal integral for $\phi(g)$ following from eq. 4.3 is:

$$\phi(g) = (\beta_0 g^2)^{\frac{-\beta_1}{2\beta_0^2}} \exp\{-\frac{1}{2}\beta_0 g^2\}(1 + O(g^2)) \ . \tag{4.7}$$

Then going to the continuum limit requires the knowledge of a definite relationship between $g$ and the lattice spacing $a$ which can be expressed in the general form as:

$$a = \frac{1}{\Lambda}\phi(g) \ , \tag{4.8}$$

where $\phi(g)$, as we have seen, is a definite function containing the information about the correct scaling behavior of $a$ when $g \to g^*$, given by eq. 4.7, and $\Lambda$ is a scale parameter.

In principle all physical quantities can be expressed in terms of $\Lambda$. One goes to the continuum limit by demanding that one (generic) observable of dimension $-d$, let us say $q_1$, remains strictly constant throughout the process of renormalization, i.e.:

$$a \equiv a(g) = [q_1]^{1/d}\phi_1(g)^{1/d} \ , \tag{4.9}$$

where $\phi_1(g)$ measures the observable $q_1$ in units of the lattice spacing. The scaling properties of the critical point involve that all the function $\phi_i(g)$ corresponding to the generic physical quantity $q_i$, of dimension $d_i$, must behave as:

$$\phi_i(g) \sim c_i[\phi(g)]^{d_i} \ , \tag{4.10}$$

for $g \to g^*$, the $c_i$ being definite constants. Then the continuum values of the observables $q_i$ are given by:

$$q_i = c_i\Lambda^{d_i} \ , \tag{4.11}$$

i.e. all physical observables are expressed in terms of the scale parameter $\Lambda$.

Then any observable might be used to establish the scale of masses and, eliminating $\Lambda$ , all other observables are expressed in terms of that one. The string tension has traditionally this role.

$\Lambda$ does not have a direct physical significance and in general it will depend on the scheme of renormalization and on the specific choice of the lattice gauge action, within the class of actions leading to the same continuum theory. $\Lambda$ will change if the renormalization scheme is modified. Then in order to relate the mainly non-perturbative lattice world to results obtained by using perturbative methods (and even, for example, different lattice results obtained by using different lattice actions) the lattice scale factor $\Lambda_L$ must be compared with the scale factors defined in the conventional scheme of perturbative renormalization as $\Lambda_{MS}$, $\Lambda_{MOM}$ and so on. Computations of ratio between lattice and perturbative scale parameters can be found in [42,43,44,45,46].

Let us summarize. Near the critical coupling $g^* = 0$ a physical observable $F$ having a dimension of $(mass)^d$ has to behave according to

$$F \sim c\Lambda_L{}^d, \Lambda_L = \frac{1}{a}\phi(g) \ . \tag{4.12}$$

where $c$ is a constant which is not computable in perturbation theory and $\phi(g)$ is given in eq. 4.7 for a pure $SU(N)$ theory. Then the expression for $\Lambda_L$ in $SU(3)$ is:

$$\Lambda_L = \frac{1}{a}\exp((-\frac{1}{2\beta_0 g^2})(\beta_0 g^2)^{-\frac{\beta_1}{2\beta_0^2}} \tag{4.13}$$

It is a very crucial step in a numerical study to examine whether the measured physical quantity obeys this scaling law. This is all the point. In a numerical simulation we will study the model for different values of the coupling constant. After checking that finite size and time effects are under control we will have to establish that indeed the measured observable follows, as a function of the coupling constant, the scaling behavior. If this is true the measured numerical constant $c$ has a physical meaning and represents the value of the physical observable we are looking at, in the continuum limit.

Then the strong coupling results valid for large values of $g$ must be extrapolated to a value of $g$ small enough that scaling toward the continuum limit is seen to take place. Also, the extrapolation should provide a self-consistency check for the theory, giving indications that no intervening critical points make the properties of the system totally different in the strong coupling domain and in scaling domain. Only in this case features such as confinement of quarks, which can easily demonstrated in the strong coupling domain, would survive the passage to the continuum limit.

The string tension has played a role of primary importance in the pioneering investigations of the scaling of lattice quantities and to verify asymptotic freedom. This quantity is regarded as the most fundamental physical quantity of lattice gauge theory and it has been continuously checked (even as a check of the computer program!) by workers in the field when they start their simulations. In particular the confining properties of the theory manifest themselves in non-vanishing string tension $\sigma$.

According to the general principles since $\sigma$ has a dimension $(mass)^2$ we expect it will scale as $\sim \Lambda_L{}^2$ on the lattice close to the weak coupling limit $g \to 0$. The aim of numerical simulation is, therefore, to extract a non perturbative quantity:

$$c = \Lambda_L/\sqrt{\sigma} \ , \tag{4.14}$$

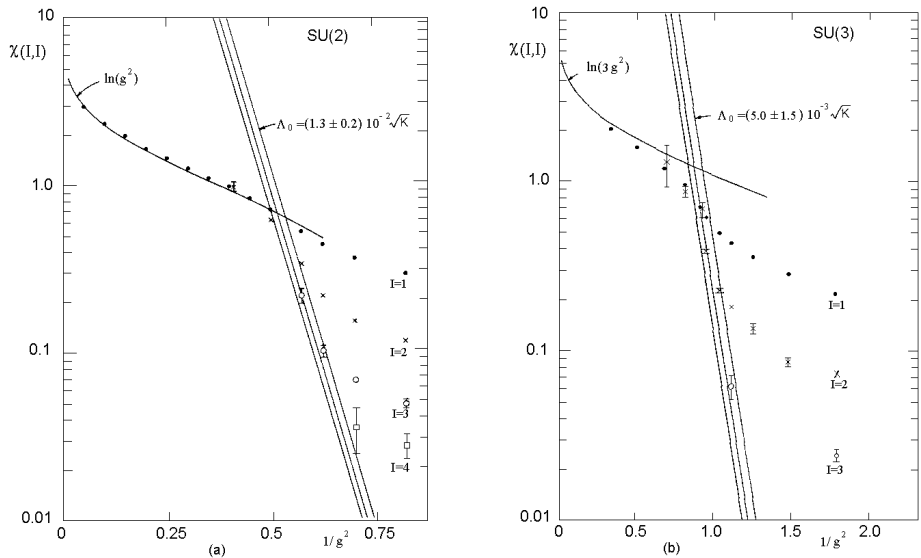

Figure 4: The quantities $\chi(I, I)$ for $SU(2)$ in (a) and for $SU(3)$ in (b), as a function of $1/g^2$. The envelope of the curves describes the string tension as a function of the coupling.

which connects the scale parameter with the string tension that controls the hadron spectrum. The string tension is a concept defined rigorously only in the absence of dynamical quarks. In their presence, it becomes energetically more favorable for large $r$ to split the string in the middle by creating a quark-antiquark pair out of vacuum.

The first investigations of the string tension were devoted to verify the scaling behavior of $\sigma(g)$ as $g \to 0$ and to use such a scaling behavior to find the relation between string tension and lattice scale parameter.

The studies in the $SU(3)$ Hamiltonian formulation[47,48] and in $SU(2)$ and $SU(3)$ Euclidean formulation[49,50] are based on the strong coupling expansion and relay on some method of extrapolation to probe the behavior of the theory as $g$ becomes small.

At the other hand the first numerical results presented in [51,14,15,16,17] are based on Monte Carlo simulations ad are independent of any extrapolation technique. The evidence for the scaling derived from this data appeared particularly impressive (we give in fig. the world famous Creutz figure, so suggestive at the time).

All this first studies are concerned with verifying the scaling behavior of $\sigma(g)$ as $g \to 0$ and with using such a scaling behavior to find the relation between the string tension and the lattice scaling parameter.

The string tension, defined as the coefficient $\sigma$ of the linearly rising part of the potential for large separations of the quark-antiquark pair, in a numerical calculation can only be extracted from relatively small $q\bar{q}$ separations (in lattice units). Hence if the scale at which this force is seen in nature is determined by the size of hadronic matter one must choose the lattice to be sufficiently coarse (i.e. the coupling $g$ sufficiently large) for hadron, represented by the $q\bar{q}$ system, to fit in it. But if the lattice is made too coarse one cannot expect to extract continuum physics, and the dimensionless string $\sigma a^2$ will not exhibit the behavior predicted by the renormalization group. Then if one chooses too small a lattice spacing (or coupling constant) then one cannot expect to see the asymptotic form of the potential on a lattice of small extent; if the lattice spacing is too large one cannot expect to see continuum physics. Thus at best one can measure the physical string tension in a narrow range of the coupling constant, picking up in this way a narrow *scaling window*. We will see at the end of this review that nowadays the situation is far more convenient, and measurements are very reliable.

We have seen that the $\beta$-function describes the relation between the bare coupling constant $g$ and the value of the cut-off, and it has a well defined meaning in the vicinity of the fixed point $g^* = 0$. As asymptotic scaling is approached the $\beta$-function helps in connecting numerical studies with perturbation theory. The $\beta$-function describes the way the bare coupling should be tuned in order to keep all the physical predictions independent of the cut-off in the continuum limit.

It should be noticed that only the two leading terms of the $\beta$-function $\beta_0$ and $\beta_1$ are universal. For large cut-off (small bare coupling values) these terms dominate and define a universal scaling behavior, the *asymptotic scaling*. Outside this region, but still in the continuum limit, the scaling behavior is described by the full, and in general, completely unknown $\beta$-function. In an $SU(N)$ gauge theory the leading corrections to this behavior are exponentially small in the inverse bare coupling constant $g^{-2}$. On the other hand the higher order corrections to the universal part of the $\beta$-function are power like and are not necessarily small in the region where the cut-off dependent corrections are already negligible. There might be sizable contributions to the $\beta$-function from non-perturbative phenomena also. Then the non perturbative $\beta$-function tells us how the lattice spacing goes to zero as $g \to 0$.

On the lattice all dimensionfull quantities like masses are measured in units of the lattice spacing $a$. In order to take the continuum limit we need to know how $a$ scales. One option is to use the 2-loop perturbative result provided it is demonstrated that this is valid at values of $g$ where the calculation (for example of string tension, glueball, etc.) are done; the other is to measure the non-perturbative $\beta$-function which quantitative structure is necessary to now since the value at which asymptotic scaling sets in is not a priori known. One should confirm also that the $\beta$-function approaches its the asymptotic form without undergoing a phase transition, providing a continuum limit with the expected properties of asymptotic freedom and confinement.

### 5. The $q - \bar{q}$ potential and improved coupling constants

When attempting to construct an effective potential to describe the interaction of quarks one is restricting his focus to heavy fermions. It is only when the mass $m$ is large that one can formulate the bound state of $q\bar{q}$ system as a non-relativistic problem, with binding energies computable from a potential via the Schrödinger equation.

General theoretical arguments give the behavior of the spin independent potential in the two extreme regions. At large separations ($r \to \infty$), confinement dominates and the physical picture is the one of a *chromo-electric flux tube*. Here the potential $V(r)$ behaves as a linear function of the distance, $V(r) \to \sigma r$. At the other end of the distance scale, i.e. at short distance, for small separations of the $q - \bar{q}$-pair, the effective potential is expected to approach the one gluon exchange result (Coulomb potential). In the continuum the perturbative form of the potential (we are dealing with an asymptotically free theory) in terms of the short-distance $q - \bar{q}$-force $\Lambda_r$ parameter[52], known by a two-loop computation, is:

$$V(r) \to -\frac{4}{3}\frac{\alpha_{q\bar{q}}(r)}{r} \ , \tag{5.1}$$

where the effective coupling $\alpha_{q\bar{q}}(r)$ behaves like $\frac{1}{\ln(r)}$ for small $r$. One has that

$$\alpha_{q\bar{q}}(r) = \frac{1}{4\pi}[\beta_0 \ln(r\Lambda_r)^{-2} + \beta_1/\beta_0 \ln\ln(r\Lambda_r)^{-2}]^{-1} \ , \tag{5.2}$$

where the usual coefficients in the perturbative expression for the $\beta$-function are given in section (4). The simplest form, in the absence of pair production processes, for the full effective potential is

$$V(r) = -\frac{\alpha_{q\bar{q}}(r)}{r} + \sigma r \ , \tag{5.3}$$

where the two quantities $\alpha_{q\bar{q}}(r)$ and $\sigma$ physically represent the scales at which an individual term begins to dominate and can be related to the charmonium or bottonium spectrum. The study of the potential can be done breaking up the separation $r$ between $q\bar{q}$ into three regions: 1) confining, characterized by the linear term at large $r$; 2) perturbative characterized by a running coupling constant; 3) intermediate $r$ values where only phenomenological forms can be tried and investigated.

The form of the ground state potential, with a Coulomb short-distance contribution and a long-range linear piece is well known and helps to account for the relative success of potential models in explaining the observed spectrum of heavy-quark mesonic states. These quark-model states do not exploit the explicit gluonic degrees of freedom present within the QCD Lagrangian, but it is possible to explore this gluonic sector through the lattice gauge theory approach in all three region of $r$. In particular the average value $W$ of an $R \times T$ rectangular Wilson loop is related to the spin independent potential $V(R)$ between static color sources on a lattice of spacing $a$ ($V(Ra) \equiv V(r)$) by:

$$V(R) = - \lim_{T \to \infty} \frac{1}{T} \ln W(R, T) \ . \tag{5.4}$$

Thus in principle the knowledge of the potential $V(r)$ allows to determine the dimensionless ratio $\sigma / \Lambda^2$ which relates the perturbative scale $\Lambda$ to a non-perturbative observable such as the string tension $\sigma$ which is, as we have seen, the coefficient of the linearly rising piece of the potential. For a proper determination of the string tension one should include in the fit the behavior of the potential at short distances. If the crossover from the perturbative Coulomb potential to the non-perturbative long distance $1/R$ term is very complex a larger than expected intermediate distance Coulomb term could be contaminating the values of the string tension.

Nevertheless the lattice calculations of the potential (where dynamical quarks are ignored) do not predict a precise functional form: one has to make a trial ansatz and use the data from these calculations to fix the unknown parameters. Then from a technical point of view the measure of the string tension is somewhat problematic, because in practice it is determined by extrapolating the heavy quark potential $V(r)$ from a distance $r$ less than 1 Fermi to large distances. Depending on which analytical form is assumed for the extrapolation, one can obtain quite different results. Then the string tension is affected by a systematic error which is not easy to control.

The measurements on the lattice of the string tension and of the potential, both related to Wilson loop measures, have *parallel histories* initiated for the potential in[53,54,55,56,57].

Here we focus our attention on the *state of the art* measurements of the potential for $SU(3)$[58,61,60,62] and for $SU(2)$[63,64,65]. We will stress the particular relevance of the key issue in taking the continuum limit of lattice QCD: the question whether the asymptotic scaling behavior holds for physical quantities as $\sqrt{\sigma} / \Lambda_{MS}$, where[42] $\Lambda_{MS} = 28.81 \Lambda_L$ for $SU(3)$ and $\Lambda_{MS} = 19.82 \Lambda_L$ for $SU(2)$.

A large body of evidence exists that the asymptotic scaling in the the 2 and 3 color QCD lattice gauge theory is not satisfied at the values of $\beta$ accessible at present (up to $\beta = 2.85$ for $SU(2)$[63] and $\beta = 6.8$ for $SU(3)$[60]. The deviation from the asymptotic scaling behavior is ascribed to finite lattice spacing effects. It is possible, however, that at least some part of the deviation is due to the presence of perturbative terms of $O(g^2)$ in the ratio $\sqrt{\sigma}/\Lambda_{\overline{MS}}$ which can be removed by a redefinition of the coupling constant. Parisi has first observed[66] (even before numerical results were obtained) that the bare lattice coupling constant $g^2 = 2N/\beta$ receives a large renormalization from gluon tadpole contributions, which makes it a poor choice as an expansion parameter. Starting from a different approach the authors of ref.[67,68] have reached similar conclusions. All together we have a systematic and quite crucial improvement of the understanding of the numerical results and expect that the scaling behavior of physical quantities will be improved with the use of different improved (if so we can say) coupling constant.

The choice done in ref.[69] is based on a mean-field version of the relation between the bare lattice coupling constant $g$ and a renormalized coupling constants such as $g_{\overline{MS}}$. The expansion parameter $g_E$ suggested by Parisi is defined introducing in the average plaquette weak coupling expansion:

$$\langle P \rangle = \sum_{n=1}^{\infty} c_n g^{2n} \ . \tag{5.5}$$

an effective coupling in terms of Monte Carlo generated plaquette expectation value:

$$g_E^2 = \frac{\langle P \rangle}{c_1} \ . \tag{5.6}$$

for which the first-order expansion is supposed to be exact. Further examples of different coupling constant schemes can be found in [58,60] and references therein.

Favourable tests in a number of examples[67,58] show a better agreement between perturbation theory and Monte Carlo results which tend to be systematically smaller than the perturbative results suggesting that a coupling defined from any of the Monte Carlo calculated quantities would yield improved predictions for the others.

In this frame the $q - \bar{q}$ potential is a natural candidate to define improved coupling constants. A phenomenological method for estimating the continuum coupling constant defined by the potential using short distance data has been proposed for $SU(2)$ in ref.[64] and extended for $SU(3)$ in ref.[63,60]. The basic idea is to calculate the coupling:

$$\alpha_{q\bar{q}}(R) = \frac{3}{4} R V(R) \ . \tag{5.7}$$

through numerical simulation and then to convert to the $\overline{MS}$ renormalization scheme warning that the conversion calculated in the one-loop approximation is only applicable when the coupling is sufficiently small, i.e. deep in the perturbative short distance regime.

In ref.[60] two approaches are presented for the determination of the QCD scale parameter $\Lambda$. First the cutoff parameter $\Lambda_L$ is calculated from the two-loop-expansion equation, following eq. (4.13).

Then the authors determine the running coupling $\alpha_{q\bar{q}}(R)$ starting from its symmetric discretization in terms of the lattice potential measurements:

$$\alpha_{q\bar{q}}(R) = \frac{3}{4} R_1 R_2 \frac{V(R_1) - V(R_2)}{R_1 - R_2} \ , \tag{5.8}$$

whit

$$R = \frac{(R_1 + R_2)}{2} \tag{5.9}$$

following the ansatz of ref.[64] for the potential:

$$V(R) = V_0 + \sigma R - e/R + f/R^2 \ . \tag{5.10}$$

The parameters $V_0, \sigma, e$ and $f$ come from a best fit. Then analyzing the $\alpha_{q\bar{q}}(R)$ data in terms of the continuum large-momentum expectation for the $SU(3)$ running coupling :

$$\alpha_{q\bar{q}}(R) = \frac{1}{4\pi} [\beta_0 \ln(r\Lambda_R)^{-2} + \beta_1/\beta_0 \ln \ln(r\Lambda_R)^{-2}]^{-1} \ . \tag{5.11}$$

and using the relations $\Lambda_R = 30.19\Lambda_L$ and $\Lambda_{\bar{MS}} = 28.81\Lambda_L$ the results are converted into the continuum renormalization ($\bar{MS}$) scheme. The two methods yield consistent values for the $\Lambda_{\bar{MS}}$-parameter

$$\Lambda_{\bar{MS}} = 0.555^{+0.019}_{-0.017} \sqrt{\sigma} \ , \tag{5.12}$$

which is in agreement with the value extrapolated from the scaling of the string tension[67,68].

This result is substantiated improving on scaling violations by replacing the bare coupling with suitable "effective" couplings: the $\beta_E^{(1)}$ coupling[66] defined by truncating the weak order expansion of the plaquette (measured in Monte Carlo simulations) after the first order term, and the $\beta_E^{(2)}$ scheme defined by truncating the same expansion after the second order term. It is demonstrated that scaling violations on the string tension can be considerably reduced allowing for a safer extrapolation of $\Lambda_L$, as a function of the different couplings, to its continuum value; the effective schemes help to decrease the uncertainty of this limit and teach that linear extrapolations can be misleading: see table 1 from first of references cited in [60].

| $\beta$ | $a\sqrt{\sigma}$ | $\sqrt{\sigma}/\Lambda_L$ | $\sqrt{\sigma}/\Lambda_L^{(1)}$ | $\sqrt{\sigma}/\Lambda_L^{(2)}$ |
|---|---|---|---|---|
| 5.7 | 0.4099 (24) | 124.7 (0.7) | 63.3 (0.4) | 55.7 (0.3) |
| 5.8 | 0.3302 (30) | 112.4 (1.0) | 63.0 (0.6) | 55.6 (0.5) |
| 5.9 | 0.2702 (37) | 102.9 (1.4) | 61.2 (0.8) | 54.3 (0.7) |
| 6.0 | 0.2265 (55) | 96.5 (2.3) | 60.0 (1.5) | 53.4 (1.3) |
| 6.2 | 0.1619 (19) | 86.4 (1.0) | 56.9 (0.7) | 50.8 (0.6) |
| 6.4 | 0.1215 (12) | 81.3 (0.8) | 55.7 (0.5) | 50.0 (0.5) |
| 6.8 | 0.0730 (12) | 76.9 (1.3) | 55.7 (0.9) | 50.4 (0.8) |
| $\infty$   Lin. | 0 | 63.6 (2.4) | 53.1(1.6) | 48.3 (1.4) |
| Log. | 0 | $54^{+/8}_{-15}$ | $53.2^{+2.6}_{-7.3}$ | $49.1^{+2.3}_{-5.9}$ |

Table 1: $\Lambda_L$ as obtained by inserting the bare lattice coupling and $\Lambda_L^{(1,2)}$ by inserting the $\beta_E^{(1,2)}$ effective couplings. Linear extrapolation to $a = 0$ results are displayed in the second to last row and logarithmic extrapolation results in the last row

One observes that the values for $\Lambda_L^{-1}$ extracted from the running of the coupling[64,63,60] tends to increase with the momentum cutoff. The approach towards asymptotic limit for different effective coupling schemes are compared[60] in Fig. 5. All that makes clear that lattice perturbation theory works better if an improved coupling constant is used.

In order to overcome the basic difficult encountered in these calculations to reach momenta $q$ greater than a few GeV it has been observed[70] that the running coupling at high energies may in principle be computed once the parameters of the theory are fixed at low energy and then is possible to make contact with the scaling region where the running coupling is determined by the perturbative renormalization group. In this frame it has been proposed to use a coupling constant[71,72,65,62,73,74]:

$$\alpha(q) = \frac{\bar{g}^2(L)}{4\pi}, \; q = 1/L \; , \tag{5.13}$$

running with the lattice size $L$. The strength of the gluon interaction at high energies can be determined by studying the scaling behavior of such an object in small and intermediate volumes. This definition is based on a recursive finite-size technique based on renormalization group arguments and is composed by two steps. The first is to match lattices with the same renormalized coupling but different lattice spacings; the second is to study of the evolution of the coupling by changing $L$ at fixed lattice spacing $a$. By combining these steps it is possible to go to the continuum limit and to follow the behavior of the renormalized constant $\bar{g}(L)$ over a large range of box sizes $L$ given in unit of the string tension $\sigma$.

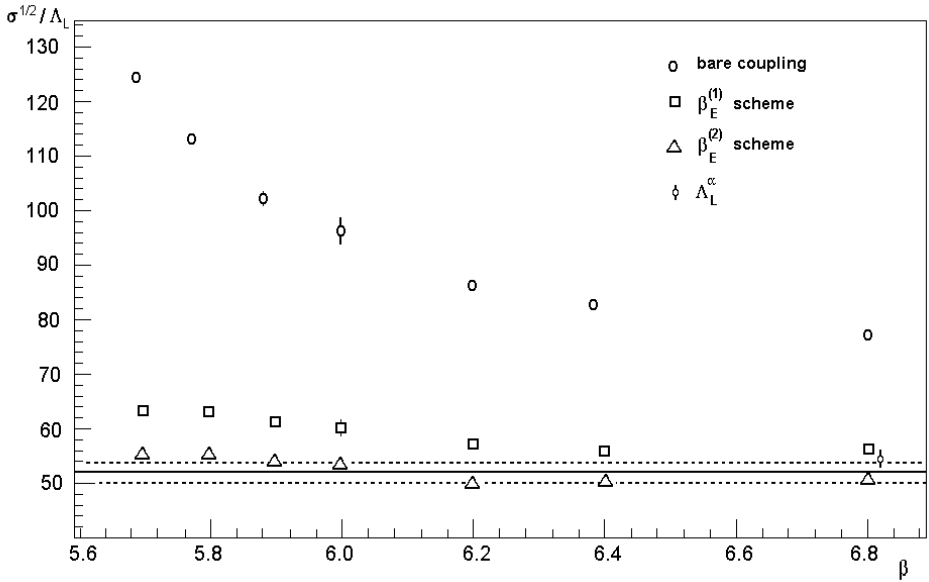

Figure 5: Violations of asymptotic scaling for different effective coupling schemes. The solid line corresponds to $\sqrt{\sigma} = 51.9^{1.6}_{1.8}\Lambda_L$.

Nevertheless the couplings defined through Wilson loop expectation values are not a perfect choice since large loops are difficult to compute numerically (the signal-to-noise ratio is exponentially decreasing with the size of the loops) and the perturbation expansion of these quantities to higher orders of coupling constant may require an unacceptable amount of work. Then $\bar{g}(L)$ is better defined through the response of the system to a constant color-electric background field, or directly by using Polyakov loops[73,74].

This non-perturbative finite-size scaling technique has been used to study the evolution of the running coupling both in $SU(2)$[65,74] and in $SU(3)$[62] gauge theories: one finds again that the perturbative scaling regime and the low-energy domain of the theories are smoothly connected with "no complicated" transition region. In particular the following relation holds within errors:

$$2\alpha_{MS}(q)|_{SU(2)} = 3\alpha_{MS}(q)|_{SU(3)} \ , \tag{5.14}$$

where $\alpha_{MS}(q)$ is determined at momenta $q$ up to $14 GeV$ with an estimated precision of a few percent; this relation is consistent with the evolution of the coupling up to 3-loop order of perturbation theory and suggests that $N\alpha_{MS}$ is only weakly dependent on the number $N$ of colours.

Some discrepancies between different numerical measurements of the running coupling can be due to non negligible non-perturbative corrections to the matching relations[62] between the different schemes, since the distances at which the comparison can be made are rather large in physical units. Cutoff effects are hardly kept under control when $\alpha_{q\bar{q}}(R)$ is determined from the heavy quark potential at distances as low as 2 or 3 lattice distances. A summary of various recent analysis for $SU(3)$ [58,10] contains results all in the range:

$$\sqrt{\sigma}/\Lambda_L = 53 \pm 10\% \ , \tag{5.15}$$

to be compared with $\sqrt{\sigma}/\Lambda_L = 80.0 \pm 1.4$ at $\beta = 6.5$ extracted by using the bare coupling. For $SU(2)$[58,59]:

$$\sqrt{\sigma}/\Lambda_L = 32.11 \pm 1.7 \tag{5.16}$$

to be compared with $\sqrt{\sigma}/\Lambda_L = 44.1 \pm .6$ at $\beta = 2.85$ extracted by using the bare coupling. These values are sufficiently far below the values extracted from bare coupling to imply that asymptotic scaling to two-loop perturbation theory is not *just around the corner* but will only be satisfied with good precision at larger $\beta$-values than those currently accessible to lattice simulation.

Nevertheless we note that the lattice methods to determine the running coupling give an accuracy comparable with that of experimental determinations of the same quantity for modest energy scales. In particular the experimental determination of the running coupling constant of $QCD$ has reached a reasonable degree of accuracy[75]. Further improvements of lattice techniques are of great interest after the discovery that the running coupling is well described by the two loop-formula down to a scale of $1 - 2 \ GeV$ till to be able to predict experimental numbers like $\alpha(M_Z)$ or $\Lambda_{\overline{MS}^{(4)}}$ as explained in ref.[69].

## 6. Methods

This section will be devoted to a brief illustration of different methods which have favored big progresses towards the study of the continuum limit and of large volume lattices in the numerical computation of the string tension.

Different methods of investigation of lattice gauge theories have been used both to measure the glueball spectrum and the string tension: in fact, as we have seen in section (3), $N_t \ K$ is the *mass* of the correlation function of two Polyakov loops and the string tension is related to the ground state of the interquark potential.

We can determine the mass of any state in an euclidean lattice calculation by measuring the exponential fall-off of correlation functions of appropriate interpolating field operators expressed by the sum over all states that couple to them:

$$\langle G(t)G(0)\rangle_c \approx \sum_{n=0} c_n \exp\left[-m_n t\right] , \qquad (6.1)$$

that, for $t \to \infty$, tends to $\exp\left[-m_0 t\right]$, where $m_0$ is the lowest mass in the channel we have selected.

To get the best estimate for the lowest energy state it is necessary to optimize the choice of the operators we use. We have an infinite choice of lattice operators with given quantum numbers. We want to get a large overlap with the physical wave function by making $c_0$ large as compared to the other coefficients. We will like to be able to measure a signal up to large distance to kill any remaining contamination originated from the excited states.

The various sources of errors to be reduced in these determinations are the following: statistical errors, finite lattice spacing errors, finite volume errors, effects of higher mass states. For excellent and more technical reviews of error analysis in lattice calculations see for example[76,164,12,58,10]

We would like to get a full understanding of the large separation asymptotic behavior of the heavy quark potential. The main problem in the the numerical calculations of the string tension $\sigma$ is that this quantity can be measured quite easily only for separations of $q\bar{q}$ of very few lattice units. We need the hadron, i.e. the $q\bar{q}$ system, to fit in the lattice, and the lattice spacing to be small enough for the discrete nature of the lattice to be irrelevant in the interaction, in order to exhibit continuum physics. But we have to go to very small values of $g^2$ in order to have a system behaving according to continuum physics.

So, the situation is not easy. For small coupling constant and small lattice spacing we do not expect to detect the correct asymptotic behavior of the potential on a too small lattice. But for a too large lattice spacing we do not expect to be dealing with a theory close to the continuum one. To check scaling one has to go to high $\beta$ values, but the lattice spacing shrinks when $g \to 0$. Operators of a fixed lattice extent become smaller in physical size and their projection over the ground state goes to zero.

Then to get best estimate, in the continuum limit, for the lowest state (which for large euclidean time dominates the correlation functions) we need good operators. Good operators means here that we want them to have a large overlap with the ground state wave function. We want to be able to measure correlation functions up to large distance, in order to control the contamination of the ground state from excited states with the same quantum number.

On the other hand the problem to extract the rate of an exponential fall-off of correlation functions of order one is a very difficult one. This is, indeed, one of the cases in which we can clearly see that a powerful computer is not really useful if we do not use the right methods to have a good control of finite volume and finite euclidean effects.

## 6.1. *MCVM*

An alternative to concentrating on small distance correlation functions is to study the correlation of Wilson loops of many types, more complex than simple plaquettes. We will extract the true long range behavior due to the lowest lying glueball state. In general one needs means of modeling the gluon flux that propagates between the static sources, thereby reducing the time propagation necessary for the effective decay of higher excited states; in the conventional approach one is modeling this flux distribution with only a straight path and consequently one has to consider correlations over larger time separations.

The idea is then to replace the limit of infinite euclidean time with the limit of infinite number of operator to pick up the ground state. In the case of the glueball spectrum this can be performed on different representations corresponding to different spin states. Then the method is to expand the glueball state in terms of certain Wilson loop operators acting on the vacuum and varying the coefficients in this expansion until the state with lowest energy is reached. In particular the spin 0 glueball states are contained in the small plaquette operator and in infinite other large operator, of different shapes. One uses this basis, which may be classified according to the discrete group of rotations, in order to maximize the projection over the fundamental state.

This procedure, named Monte Carlo Variational Method (MCVM) was proposed by Wilson[77], first applied for the glueball masses in [78,79] and then used in [80,81,82] and for the potential function of a gluonic string with fixed ends in [54].

Let us explain Wilson's proposal applied on the lattice string states, with some details.

One studies the gluonic field between a static quark and antiquark by considering, in principle, the traces of an infinite number of Wilson loops of the type shown in fig. 6. The time directed links represent static charges, while the space directed paths $P_i$ and $P_j$ create and annihilate gluons from the vacuum.

The correlations $C_{ij}(R,T)$ are formed by parallel transporting the path $P_i$ through time $T$ and taking its trace with the path $P_j$. Their dependence on euclidean time is given by the appropriate eigenvalues in the transfer matrix formalism[8].

A matrix variational technique will give, for each symmetry of the gluon field, the linear combination of $C_{ij}$'s which decreases least rapidly allowing to extract the largest eigenvalues of the transfer matrix from relatively small values of $T$. For example with $T = a, 2a, 3a$, and 3 independent path combinations it would be possible to extract 4 eigenvalues, whereas a direct study of rectangular loops of length up to $T = 8a$ would be needed to obtain the same information in principle[55]. Since loops with large values of $T$ have small average values, the errors are relatively large and so accurate numerical calculations of their averages are very time consuming. Conversely the MCVM measures a large number of Wilson loops of different shapes (the overlaps between path $P_i$ at $T = 0$ and path $P_i$ at $T$) at relatively small $T$ values. In this way it enables a more accurate extraction of the ground state and then of the interquark potential and of the string tension.

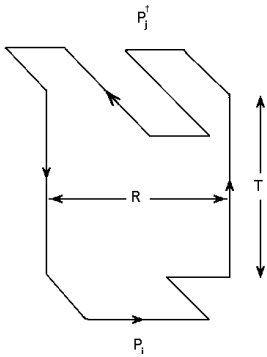

Figure 6: The Wilson loop giving the correlation $C_{ij}(R,T)$ between the spatial paths $P_i$ and $P_j$ separated by time $T$ as illustrated in ref.[54].

We note that if the theory has a positive definite transfer matrix the estimate we get at finite time for the ground state is an *upper bound* to the true asymptotic value.

The main problem with this method is due to the statistical nature of our knowledge of how much we need of each different operator. The minimization procedure, difficult if the error on the measured operators is not very small, can in some cases result quite ineffective.

## 6.2. *The source methods*

The search of statistical improvements needed in order to get sensible measurements of long distance correlation functions has been one of the main developments of lattice gauge theories. One has started to deal with non local objects, most likely spread out over a complete time slice and a variational calculation will need too many loops; then it is necessary to enhance the signal at large separations.

The use of a source method has been proposed, in different forms, in ref.
[83,84,85,86,87]

In order to study correlation functions in a medium (the lattice filled up with gluons for example) it is possible to perturb the system and to look at the response to such a perturbation. This is analogous to what one does for studying the heat conduction properties of a given material.

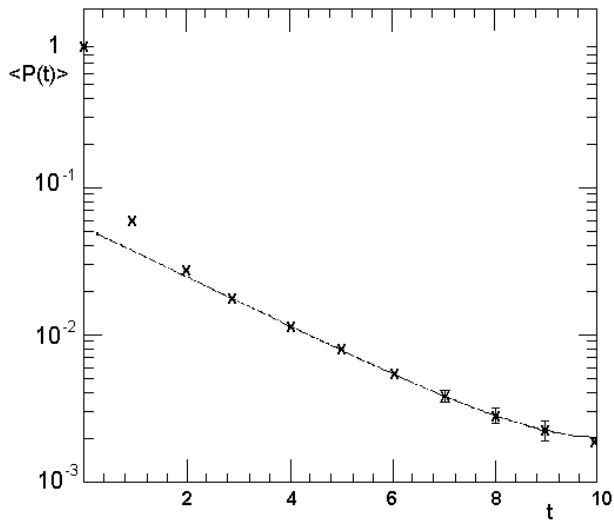

Figure 7: The response of Polyakov loop operator, $\langle P(t)\rangle$, to a cold wall source, from ref. [89].

In particular one can look at the decay of the signal coming from a *cold wall*[84,83]. The source method, as applied in ref.[87], is based on the study of the response of the expectation value of an operator on several space slices at times $T = 2, 3, ..., L$ setting all spatial links at time equal one to the identity.

Thanks to the effect of the wall which generates a strong signal one can measure the exponential decay for larger distances than using loop-loop correlation functions (see fig.7). This is the reason for the use of a source in computations of the string tension.

The difference from other methods is that in the case of the source the transfer matrix is not more positive definite and the value estimated for the ground state at finite time is not an upper bound to the true value anymore.

Moreover, in the presence of the source, both the procedure of smearing (discussed later) the gauge field in space[87] and the variational method[89] turn out to be useful.

### 6.3.  *The difference method and Langevin equation*

In the Monte Carlo approach correlation functions are obtained as expectation values of operators averaging over a suitable number of uncorrelated configurations. Since large distance correlation functions are small, and they are normally determined by measuring terms of order one, they are dramatically affected by statistical fluctuations. For this reason a large number of configuration is needed to obtain a satisfactory precision in correlation functions measurements.

Then the idea of perturbing the system far away from equilibrium in a limited space-time region and measuring the decay of correlations from such a zone can be pushed far away by using specific properties of the dynamical updating algorithm. The selection of such an algorithm can play a crucial role in helping to get more accurate results.

The method based on this idea looks at differences of the values of the same quantity on two configurations. If we are able to keep the two configurations very close each other, the contribution of the statistical fluctuations affecting the two configurations cancels in the difference.

To describe the *difference method*[83,85,86,87] one considers two gauge systems $K$ and $K'(i)$ on two lattices of identical size $M^3 \times L$; $K$ is kept at inverse square coupling $\beta$; $K'(i)$ has a "time slice" $i$ $(i = 1, 2, 3...L)$ where the inverse square coupling takes the value $\beta + \delta\beta$.

We define $E(j)$ as the average energy of the time slice $j$ for the system $K$, and $W(i, j)$ as the average energy of time slice $j$ for the perturbed system $K'(i)$. Then we can show[83], using Wilson action, that the connected correlation function at distance $d$ of the energy operator $C(d)$ is given by:

$$C(d) = \frac{E(j) - W(i, j)}{\delta\beta} , \qquad (6.2)$$

for any $i, j$ such that $d = |j - i|$.

It is easy to see that if we generate independent configurations by using a normal Monte Carlo simulation we are not in good conditions to compute correlations functions by this technique. In fact, in order to avoid noisy correlations the two sets of configurations which produce the averages $K$ and $K'$ must be *similar* (we are forgetting for a moment about gauge invariance). This cannot be achieved by a Monte Carlo algorithm, like a Metropolis updating, which does not give a continuous trajectory in the phase space; it is an 'yes or no' procedure and the evolution trajectories of the two systems can just choose if being identical or completely different. In fact a given change of a gauge variable can be accepted at $\beta$ while the corresponding at $\beta + \delta\beta$ is discarded. Then we need a method to generate configurations continuous in the $\beta$-variable and the increase in accuracy over the Monte Carlo method is achieved just by allowing for a coherent cancellation of statistical errors between two highly correlated stochastic processes.

Both Langevin[90,91,85,86,83,92] update scheme and an heat bath method made continuous[87] can generate sets of *similar* configurations. In the first scheme, if $S$ is the action, the gauge field configurations can be obtained as solution of the following Langevin equation:

$$\dot{U}_L(t) = \frac{\delta S}{\delta U_L} + \eta_L(t) \; , \qquad (6.3)$$

where $\eta_L(t)$ is a gaussian noise with autocorrelation:

$$\langle \eta_L(t)\eta_{L'}(t')\rangle = \delta_{LL'}\delta(t - t') \; . \qquad (6.4)$$

We note that two different *times* come into play: one is the physical euclidean time, one of the four dimension of the lattice, completely homogeneous to the other 3 space dimensions. The other, a computer time, is the evolution time of the unphysical dynamics of the differential equation used to formulate Langevin algorithm for lattice gauge theory. After a certain amount of this computer time, the system will be representative of the probability Boltzmann distribution; therefore the Langevin equation provides a way to generate this distribution analytically and numerically.

One has to be careful in noticing that the discretized version of Langevin equation requires the introduction of new parameters, that will have to be tuned in order to get good performances.

On the other hand it is known that very efficient updating algorithms are the heat bath ones[93,51,14,88]. It is very easy to make them to satisfy the requirement of a continuum trajectory; for the $SU(3)$ gauge theory the method proposed in ref.[88] has been discussed and applied in[87]. In particular this second method exhibits two relevant improvements with respect to the numerical solution of the Langevin equation: it is about ten times faster in reaching thermal equilibrium and does not need arbitrary parameters arising from the discretized version of the equation.

Besides we note that gauge invariance effects increase the noise on the system and then gauge fixing or something equivalent must be imposed both in Langevin and heat bath updating algorithms[83,87]. In particular the main disadvantage of this second method is the increasing of the noise for large upgrading *times*. An increase in the noise by a factor 10 or more occurs in a few iterations, then the two configurations become quite different after the onset of the strong noise. Moreover, using an heat bath algorithm, it is convenient to measure correlation functions for operators which are a functional of the field smeared in space.

### 6.4.   *The DLR and multihit methods*

Another strategy for reducing the statistical errors in the evaluation of Wilson or Polyakov loops for large lattices is to replace the original observable with its local average. One evaluates the thermal average of a link with respect to its nearest neighbor links. The method proposed in ref.[22,94] is based on the identity derived by Callen [95], a particular case of the DLR equation[96]. The idea is very simple: if two observables $A$ and $B$, with statistical errors respectively $\langle A^2 \rangle_c$ and $\langle B^2 \rangle_c$, are such that:

$$\langle A \rangle = \langle B \rangle \text{ but } \langle A^2 \rangle_c \gg \langle B^2 \rangle_c \ , \tag{6.5}$$

it is obviously much better to measure $B$ instead of $A$. In ref. [22] the $SU(3)$ gauge link variable $U$ is replaced in the Polyakov line with

$$\overline{U} = \int dU \ U \ \exp{-\beta H} / \int dU \exp{-\beta H} \ , \tag{6.6}$$

where the integration is done over the links neighboring the original one.

A heat bath method is used to evaluate $\overline{U}$ numerically. In particular $N = 14$ heated hits (a *multi-hit* method [94,56,97]) have been done and it has been seen that the statistical fluctuations decrease with $N$ down to a plateau value which is reached for $N \approx O(10)$.

### 6.5. *Smearings and fuzziness*

The "first generation" of string tension calculations used small loops, such as the plaquette. It is known that for local operators the noise-to-signal ratio diverges as the continuum limit is approached (as discussed for example in ref.[98]).

In particular to check the scaling one has to compute perturbatively the dependence over $\beta$ at the critical point of the theory. Then it is necessary to go to higher and higher $\beta$ values despite the fact that the lattice spacing shrinks when $g^2 \rightarrow 0$. Operators of a fixed lattice extent become smaller in physical size, and their projection over the ground state goes to zero. To improve one must resort to non local operators.

Actually several methods[13,12,58] are known to work to kill the unphysical short wavelength fluctuations. In particular two iterative methods : the *smearing*[87,102,103,104,105] and the *fuzzy loops*[27,99,101,100], have been introduced.

The smearing procedure as originally proposed for $SU(3)$ in [87] consists in the construction of correlation functions for operators which are a functional of the field smeared in space and not in time. To this end it is possible to construct a block gauge field with a simple gauge invariant procedure.

For each link of a generated configuration we consider the product of the other three link variables defining with it a plaquette and we sum these product over 4 choices which define plaquettes orthogonal to the time axis; the resulting matrix, projected on the gauge group, will be the new link variable; a smart description[12] of the procedure is in fig. 8.

The smearing coefficient $\epsilon$ is chosen by optimizing the performance of the method[102].

This procedure can be iterated (see fig.9) to obtain a field more and more smeared in space at fixed time. One can evaluate a series of operators in which short distance fluctuations are more and more suppressed as the gauge field on the link represents a smeared average over larger and larger neighborhoods. Then one can have a set of operators, to implement a variational calculation, allowing to evaluate if a given euclidean time distance is asymptotic; otherwise, as the amplitude of different exponential contributions depend on the operator, the ground state mass estimate from that given time distance can depend over the choice of the operator.

Figure 8: Description of Ape's algorithm for smearing from ref. [12].

In momentum space the effect of the smearing is equivalent to the application of the factor $\exp -k^2$ (as it is shown in ref.[102] for the example of a scalar field). This consideration makes more explicit the role of suppressor of short wavelength fluctuations that the smearing has.

The fuzzy procedure[12] is depicted in fig.10

In principle the $\alpha_i$ coefficients are tunable even if $\alpha_1 = \alpha_2$ is chosen in ref. [100] to normalize the fuzzy link back into $SU(3)$. We note that the fuzzy procedure is inspired by the Monte Carlo renormalization group (which we will discuss later) methods involving factor-of-two blocking.

Both fuzzy blocking and smearing procedures are performed only in the spacelike directions because unambiguous identification of the masses relies, through the transfer matrix, on operators with have support on one time slice.

A consequence of fuzzy technique is that *links paths* grow exponentially during the iterations and there are $2^3$ times fewer fuzzy links than original links. Then, after iteration, a simple loop of fuzzy links is a complicated linear combination of loops of original links. Finally elementary loops of fuzzy links are quite non local when expressed in terms of the original links.

A slight variation of the APE-style smearing proposing[106] an alternative way of constructing operators improves the signal for a fixed number of smearing steps. Nevertheless, as $a \to 0$, all these methods become critically slowed down.

An accurate test on efficiency of various smearing algorithms and a comparison with the cold wall source method are performed in ref.[106].

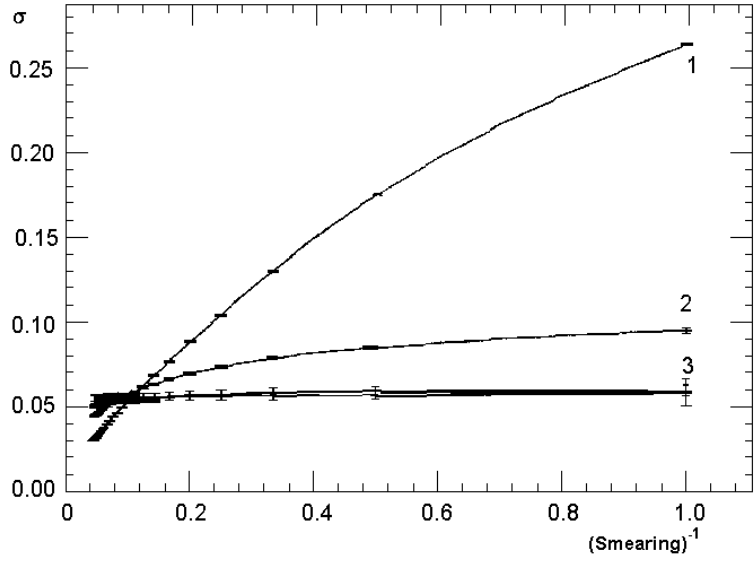

Figure 9: The string tension $\sigma$ estimated at different times versus the operator plotted as a function of the inverse smearing number from ref.[102].

Figure 10: Description of Teper's algorithm for fuzzy links from ref. [12].

### 6.6. *MCRG method and Scaling violations*

The Monte Carlo Renormalization Group method is known to give accurate and reliable information about the critical properties of statistical systems[107] [108] [109] [110] and lattice gauge theories [111,112,113]. It is a combination of the principles of Monte Carlo simulation[114] with those of real space renormalization group [115,116]. Here this method is analyzed only in connection to the study of the $\beta$-function related to the measures of the string tension and its scaling violations.

It is in general an open problem how big a lattice in a gauge theory simulation should be to do physics really near the continuum. The determination of the $\beta$-function controlling the relation between the coupling constant and the cut-off is very relevant here. One would like for example to know where the $\beta$-function changes from being representative of the non-perturbative to the perturbative theory.

Since a confirmation of the approach to the continuum limit has basic importance in lattice QCD, systematic scaling analysis at large values of $\beta$ are inevitably required. In particular the measurement of the $\beta$-function is crucial in lattice gauge theories. It is very difficult to establish the connection to the continuum theory without it. The $\beta$-function describes how the bare coupling should be tuned in order to keep all the physical predictions independent of the cut-off in the continuum limit. The $\beta$-function is unique in this sense but not universal: it is different in different lattice formulation. In particular the $\beta$-function depends on the lattice action chosen. Only the two leading terms in its perturbative expansion are universal:

$$\beta(g) = -\beta_0 g^3 - \beta_1 g^5 + O(g^7) \ . \tag{6.7}$$

as given in eq. (4.5) For large cut-offs (small bare coupling values) these terms dominate and define a universal scaling behavior, the *asymptotic scaling*. Outside this region, but still in the continuum limit, the scaling behavior is described by the full, and in general unknown $\beta$-function. This function gives the way asymptotic scaling is approached, connects numerical studies with perturbation theory, reveals the existence of possible phase transitions.

In the MCRG approach, not the $\beta$-function itself , but a related quantity $\Delta\beta = \Delta\beta(\beta)$ is determined, which gives the change of the coupling

$$\beta \to \beta - \Delta\beta(\beta) \ , \tag{6.8}$$

when the (dimensionless) correlation length (or the cut-off) is decreased by a factor of $b$. Here $b$ is the basic change of scale in a single renormalization group (RG) step ($b = 2$ in the following). At the couplings $\beta$ and $\beta\prime = \beta - \Delta\beta(\beta)$ the model has identical long-distance properties, only the (dimensionless) correlation length $\xi$ differs by a factor of 2.

A central part of any MCGR study is the block transformation (BT) which calculates the block fields on the next coarser block lattice from averaging link variables over appropriate regions of the current lattice. The system under consideration is divided into *blocks* and a smaller number of block variables are defined by averaging in some fashion over the original *site* variables. By studying the way in which the site hamiltonian (or lagrangian for lattice gauge theories) "flows" into the block renormalized hamiltonian one can determine the critical properties of the system.

The most important properties of a BT are:

1. The partition function of the two systems (unblocked and blocked) is unchanged but the correlation length of the block variables in units of the block lattice spacing is one-half that of the original variables in units of the original lattice spacing. This rescaling of lengths is the key to any RG transformation.

2. The interaction between the *blocks* is described by a new action which in general will contain all kinds of interactions.

Here we have used the crucial assumption that a fixed point exists and that it is short ranged. Renormalization generates an infinite number of couplings, but we assume that a small number of short range points is enough to describe the system at a given scale and the long distance physics is preserved.

The new action generated through iteration of the RG transformation can be considered as a point in the multidimensional space of the different couplings constants and a BT transformation is expected to have a fixed point somewhere in the $\beta = \infty$ hyperplane of this multidimensional space. A single renormalized trajectory (RT) starts form this point (see fig.11 from ref.[117]).

Both the positions of the fixed point and of the RT are not universal and depend on the BT. We start with the standard action at a given $\beta$ value (for large $\beta$). The effective actions obtained after a few BT will move close to the RT. The same will happen if we start at some other coupling values $\beta'$: by tuning $\beta'$ the points of this second sequence, which lie on the RT, can reach the corresponding points of the first sequence but one step behind.

Different procedures have been proposed to implement the MCRG. The first one originally proposed by Wilson[111] is to find an action close to the RT of a given BT. A different strategy is to look for an *improved* BT whose fixed point and RT lie close to the standard action[117,97,121].

An other method[122] is based on the Schwinger-Dyson equations, providing the block renormalized couplings and the RT for any value of the initial couplings, also far from the critical values. An accurate review of these methods is in ref.[164].

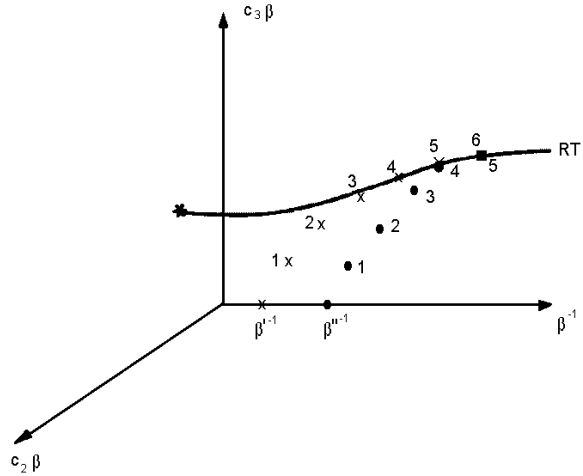

Figure 11: The RT represented in the hyperplane of the multidimensional coupling constant space. The fixed point lies in the $\beta = \infty$ plane and the standard action corresponds to the points of the $c_2 = c_3 = ... = 0$ axes

In this section we will describe a generally applicable renormalization transformation (BT) for analyzing lattice gauge theories as presented originally in ref.[113]. Others BT are used in the literature: in fact many different BT can be used to analyze a given model and their comparison is useful to check block transformation universality[164,119] and efficiency.

The use of real-space RG in lattice gauge theories has proven more difficult than for the corresponding spin systems because of the high symmetry of a gauge theory. Most simple definitions of *block spins* fail to preserve the gauge symmetry and can only be used after some form of gauge fixing.

The transformation introduced in[113] avoids gauge fixing either globally or within specified blocks[111], is directly applicable to arbitrary lattice gauge theories and the renormalized Hamiltonians (actions) retain the full gauge symmetry of the original model.

This BT is illustrated[113], for simplicity, on a two-dimensional lattice in fig.12.

The lines represent the $U_{ij}$ gauge variables and the intersections are the sites of the original lattice. For a transformation with scale factor 2 only the circled sites remain in the new renormalized lattice. The renormalized gauge fields connecting two sites of the *blocked* lattice are constructed from the operator products along the paths marked $A$, $B$, and $C$ as shown in fig.12

To complete the description of scaling analysis by MCRG methods, performed by several groups[117,97,120,123,124,125,126] we recall two methods for cal-culating the non perturbative $\beta$-function directly: the loop ratio method[118,19][117,97,120] and the Wilson's two lattices method[111,112,113,120,123,126]. The second method, designed for models with one relevant direction in coupling parameter space, is used in the one of the most recent scaling study[126] of quenched pure gauge lattice QCD.

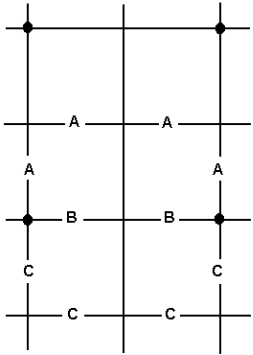

Figure 12: Diagram of the renormalization transformation for lattice gauge theories as proposed in ref. [113] sketched in two dimensions. Operator products are defined along each of the paths marked A, B, and C.

This method is based on the blocking which is repeatedly performed down to $2^4$ for two lattices, one of size $L$ (at $\beta$) and the other of size $L/2$ (at $\beta - \Delta\beta$). To match long range physical contents on both lattices, a set of Wilson loops on one blocked lattice is compared with the corresponding one on the other blocked lattice.

The expectation values from the two simulations are then compared on the same size lattices, i.e. the ones from the larger starting lattice $L$ blocked one more time than those from the smaller lattice. The test for convergence of the two systems is that the expectations values should match simultaneously at the last few levels. For early matching between blocking trajectories, the blocking transformation is controlled by a parameter $q$ which governs the size of Gaussian fluctuations around the maximal $SU(3)$ projection of the block link variables[123].

The coupling shift $\Delta\beta$ is determined at the value of $q$ where the mismatch of the two sets of Wilson loops is minimum; planar $1 \times 1$ and $1 \times 2$ Wilson loops, non-planar 6-link ("twist" and "chair") and 8-link ("sofa") loops are measured on blocked lattices (fig. 13)

In the frame of the MCRGM different updating methods have been used: the heat-bath[14] and the over-relaxed pseudo heat-bath algorithm [127].

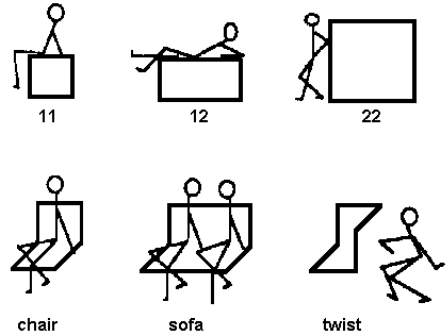

Figure 13: Loops used in ref.[126].

## 7. The use of different actions

An important question raised just after the first measurements of the string tension was how close were the measured values to the real, continuum content of the model considered.

Wilson's point of view[2] that confinement persists for arbitrary small values of the bare charge seemed to be confirmed by the first results. A possible useful internal check was to compare results obtained by using different lattice actions in the scheme of the universality. A possibility to reduce lattice artifacts using different actions is based on the idea that for a large class of actions the physical predictions are universal while the cut-off effects and in general the lattice artifacts are not. No matter what the form of the lattice action is once the gauge group is fixed the continuum physics should be defined. In particular the string tension as measured by Monte Carlo methods should behave for small enough bare charge in a universal way, up to a multiplicative constant determined by the action in question. The change of the lattice action corresponds to modify the scheme of regularization but the continuum limit should not be affected.

The corresponding ($\Lambda$) scale parameters can be exactly connected by a perturbative calculation (in the $g \to 0$ limit) [141,135,147,136] and it is possible to check if the Monte Carlo simulations follow such predictions.

The results [137,135,78,138 153,139,157,140,154,106] referring to different actions, (for example the Manton's action[128], the heat kernel action [129,130,131,132] and the mixed actions consisting of terms belonging to fundamental, adjoint and others representations[133,106]), show that the scaling form of the string tension is reached in different coupling region for the various actions. In ref.[133] it is discussed how a modification of the Wilson action can introduce a first-order phase transition between the strong and weak-coupling regime in $SU(2)$ lattice gauge theory; nevertheless this transition is not deconfining because it can be continued around in a larger coupling space. The hypothesis that *more structure* could be present in the string tension was investigated also in[141].

A test of the universality and scaling for the measure of the string tension whit Manton's, Symanzik's tree-level (discussed later) and Wilson's action in $SU(2)$[142] shows that significant violations of scaling are similar for the three actions and suggests that the reason for this violations is common to a whole subclass of lattice discretizations and could be associated with the underlying cubic lattice substructure. Manton's and Symanzik's actions violate asymptotic scaling in the same direction as Wilson action but significantly more weakly; nevertheless such an improvement is payed with the difficulties arising from the violation of positivity characterizing improved actions[142].

So the use of the improved actions had a limited effect on actual calculations until recently. This has also been because of the poor performance of bare perturbation theory. A description of different methods of improving the action, with their advantages and disadvantages can be found in ref.[164] and the state of art of these procedures in[143,144] and references therein.

## 7.1. *Symanzik's improved actions*

In the frame of finding an improved action in lattice gauge theories to reduce the effect of operators that lead to scaling violations the Symanzik's improvement program[145,146,147,148,149,150] assumes a particular relevance.

This program is based on building a systematic reduction of the cutoff dependence of physical quantities, diminishing the corrections to continuum theory stemming from finite lattice spacing. If a theory is regularized by introducing a lattice spacing $a$, when $a \to 0$ the regularized theory should go to the continuum theory; however, when $a$ is different from zero the two theories are different. Symanzik's improvement program consists in choosing the action on the lattice in such a way that the difference between the two theories is as small as possible.

Symanzik's scheme specifies in a precise way how to quantify the difference between the lattice regularized theory and the continuum theory and shows that the improved action, defined following the criterion he has proposed, can be computed in perturbation theory or measured in conceptually simple Monte Carlo experiments.

As noticed in section (4) any of the physical quantities calculated in a lattice regularized version of the gauge theory with perfect scaling will obey the renormalization group equation (4.1).

For a generic action this will however not possible and, according to perturbation theory, the best one can achieve is approximate scaling near the continuum limit, i.e. for $a \to 0$:

$$\left( -a\frac{\partial}{\partial a} + \beta(g)\frac{\partial}{\partial g} \right)[\text{physical quantity}] = O[a^2(\ln a)] \tag{7.1}$$

The suggestion of Symanzik's improvement program is to reduce the scaling violations to terms at least of order $O(a^4(\ln a))$ to all orders of perturbation theory including in the lattice action suitable chosen irrelevant terms. Actually it is important to note that not only spectral quantities can be improved, but also the (properly normalized) $n$-point functions of the fundamental fields in momentum space.

In ref.[151] on can find a review of the theoretical basis, the possible applications of Symanzik's improvement program and a discussion of his possible extensions to be compatible with the positivity and symmetry of the transfer matrix. In fact the Osterwalder-Schrader positivity condition is crucial in numerical simulations: it is at the heart of the MCV method discussed in section (6.1).

For $4d$ $SU(N)$ lattice gauge theories the Symanzik's tree-level improved action (TIA) has been determined[146,147,148,149]. A motivated ansatz[147] for the improved action includes Wilson loops up to length 6. The result[149,148] is a function also of the planar rectangular double plaquettes of size $1 \times 2$.

Different Monte Carlo studies of string tension have been carried out using Symanzik improved actions[152,153,154,142]. In particular we refer to studies of Creutz ratios up to $\chi(4,4)$[154]: an improvement of previous string tension estimates[14,16] is allowed and it is also shown that Creutz ratios $\chi(I,J)$ are not stable under increasing $(I,J)$ from $(3,3)$ to $(4,4)$. It is this a clear tendency of the estimated string tension to decrease as already noted in that period by a potential analysis[157] and a Polyakov loop measurement[155,156].

We also note the relevance of a consistent string tension measurements involving both the improved action and corresponding improved Creutz ratios which expression at tree level order is[148]:

$$\chi^I(I.J) = \sum_{m,n} C_{m,n}^I \ln[W(I+m, J+n)] . \tag{7.2}$$

where the improved $C_{m,n}^I$ coefficients and the corresponding ones of the standard Creutz ratios can be found in ref.[154].

Scaling of the improved string tension sets on at a smaller correlation length than for the standard plaquette action but scaling windows, for similar lattice sizes, look similar and therefore, from a practical point of view, the improvement is modest. An important consistency check is, of course, the observation that the improved definitions[148] of Creutz ratio work well.

In a following study[142] calculations for larger lattices show that the string tension follows 2-loop perturbative theory more closely in the case of the TIA and Manton's actions than in the case of the standard plaquette action;

still there is a warning on the difficulties caused by the violation of positivity of improved actions. In particular in ref. [142] it is showed that in the case of the Wilson block-spin-improved action[111] the violations of positivity make impossible to extract accurate masses at small $\beta$ values. One has to be careful in giving a final word about the program based on understanding continuum physics at smaller $\beta$ values using *improved* actions.

## 8. String picture

An alternative way to parameterize the lattice data in the large $r$ region is to assume a simple long distance picture of QCD, the one of a chromo-electric flux confined to a tube. The color flux joining a pair of quarks in the confining phase is concentrated inside this tube of small but finite thickness. The possibility of describing the long-distance dynamics of gauge theories in the confining phase by an effective string theory[158,159,160] is based on the very intuitive assumption that the color flux connecting a pair of distant quarks is squeezed, in the confining phase, within a narrow flux tube (string) carrying constant energy density; as a consequence, the energy of the system increases with the separation of the quarks.

It is generally believed that this thin tube behaves like a vibrating string when the quarks are pulled very far apart. This is also supported by the strong coupling expansion of the lattice gauge theories which can be formulated as a sum of weighted random surfaces with quark lines as boundary. The string picture of confinement in the strong coupling approximation within the framework of the Hamiltonian formulation was discussed originally by Kogut and Susskind[161].

The fluctuating string can be taken as the description of the color flux between colored sources: since the string model appears to be reasonable it is interesting to quantify the nature of this string. One can study the modes of a scalar gaussian string[162] to derive which terms contribute to the potential. In particular, in lattice calculations, static color sources can be used, so that an appropriate method to explore the string model is a study of the long range static potential $V(R)$.

Nevertheless the action describing the effective string in the continuum limit is substantially unknown. The simplest possible assumption is that the action of the effective string is described by the Nambu-Goto string[162,163] in terms of $(d-2)$ free bosonic fields associated to the transverse displacements of the string. The leading behavior of a $R \times T$ Wilson loop is[164]:

$$-\ln W(R,T) = \sigma RT + p(R+T) + c - (d-2)\{\frac{\pi T}{24R} + \frac{\ln R}{4} + \frac{1}{2}\sum_{n=1}^{\infty}\ln(1-e^{\frac{2n\pi T}{R}})\} \ .$$

$$(8.1)$$

The last term is universal and only depends on the number of transverse dimensions $(d-2)$. The parameters $\sigma, p, c$ depend on $g$.

Different kinds of difficulties are present in applying the bosonic string outside the critical space-time dimension of 26. Modifications[165,166] introduced to transform the Nambu-Goto action into a consistent theory make it very difficult to evaluate physical observables which should be the final goal of the effective string picture of the gauge theories. In effect, considering only the infrared behavior of the interquark potential, the effective string is asymptotically described by a two dimensional conformal field theory formed by $(d-2)$ massless free bosonic fields which is the infrared limit of the Nambu-Goto action called the *free bosonic string*.

Different schemes are proposed to overcome the too drastic approximations given by the free bosonic string. A simple modification of this picture[167,168] consists in a suitable compactification of the boson fields. It fulfills the constraints dictated by the gauge system[169] which cannot be obeyed by the free bosonic string and fits well the numerical simulations of the gauge systems in three and four space-time dimensions and with various gauge groups. The effective string picture arising in this way coincides with that proposed previously[170,171] in order to fit accurately the numerical data on the expectation value of the Wilson loops for various gauge systems in three and four space-time dimensions and based on a different argument[169].

A numerical exploration of the string picture studying the long range static potentials between static fundamental color sources corresponding to the ground state and excited states has been performed in ref.[57]. The prediction of this picture is that the ground state potential between static quarks consists of a term linear in the separation $R$ of the sources and of corrections to this term originating from the zero point motion of the quantized string. The leading correction is proportional to $1/R$ and the constant of proportionality has the universal value $\pi/12$[163] for a general class of bosonic string theories. Considering the Nambu string[162,163] with fixed ends in four dimensions the corresponding potential for the different string states is given by the formula:

$$V_N = (K^2 R^2 - \pi K/6 + 2\pi N K)^{1/2}, \ \ N = 0, 1, 2, 3, ... \qquad (8.2)$$

The agreement between predictions of this picture and lattice simulations is good at relatively large values of $R$[57] where the width of the flux tube is smaller than its length and the string model is expected to apply. Problems and discrepancies arise for small $R$.

However the quantum nature of the string is only revealed by study of the string fluctuation term and more subtle tests involve the excitations modes of the string which have energy $\pi/R$ per mode at large $R$. Then the string picture is clearly a useful guide to the excitation energies at large $R$.

As it is well known[163] the string fluctuation component in $V(R)$ for a bosonic string behaves as $\pi/(12R)$ at large $R$ and a keen test concerns just this self-energy term which is difficult to separate unambiguously from the contribution to the potential of an effective Coulomb term dominated by one-gluon exchange also behaving as $1/R$. Suggestions have been made that other self-energy expressions are appropriate for different string models. The Dirac string described above[167,168] (involving the Coulomb piece fixed by lowest order perturbation theory) leads to a self-energy term four times smaller.

So it would be useful to determine the self-energy contribution accurately from lattice calculations. The difficulty to separate this term from the Coulombian component lead to the proposal[172] to explore 3-dimensional gauge theories where the Coulomb and string fluctuation effects are clearly different ($logR$ and $1/R$ respectively). Attempts to separate effects due to gluon exchange (Coulomb) and string fluctuations (confining force) from an analysis of the spin orbit potentials[173] have been performed [174] but more accurate results are needed to confirm the validity of this approach.

Another way is to study closed strings via states of electric flux which encircle the periodic spatial boundary conditions, called *torelons*, of size $L$. These are systems which energy $E(L)$ will again have a string tension component $KL$, but where the string fluctuation term behaving as $\pi/(3L)$ for a bosonic string is unaffected by any Coulomb component; so a search for the self-energy is propitious. To extract the string fluctuation component one should consider long strings so that the string model is most likely to apply; $L \times S^2 \times T$ lattices with both $S$ and $T$ large have been analyzed[175].

Accurate measurements of the energy of the ground state torelon $E(L)$ show confinement generated through a color flux tube having energy which increases with his length: the existence of a constant confining force (the string tension $K$ ) is confirmed in a non-abelian $SU(2)$ color gauge theory. A careful study[176] of $E(L)$ of the torelons can expose the possible self-energy contributions in a hadronic string. Following ref.[57] the parameterization is:

$$E(L)^2 = (KL)^2 - \frac{2fK\pi}{3} \ ,$$ (8.3)

which for large L becomes:

$$E(L)/L = K - f\pi/3L^2 \ .$$ (8.4)

The comparison[176] of the fits with the data is illustrated in fig.14.

The slope in the figure is directly proportional to $f$; the continuous line shows the fit which gives strong evidence for a term in $E(L)$ with $L^{-1}$ behavior at large $L$. This term is expected from a string self-energy contribution and the coefficient $f$, compared with that of simple string models, is found completely consistent with the bosonic string value of $f = 1$ and inconsistent with $f = 1/4$, the Dirac string value. For the conventional string model there is no discrepancy between the string tension extracted from potentials and torelons. This results are valid for pure gauge $SU(2)$ color fields and similar conclusions can be expected for $SU(3)$.

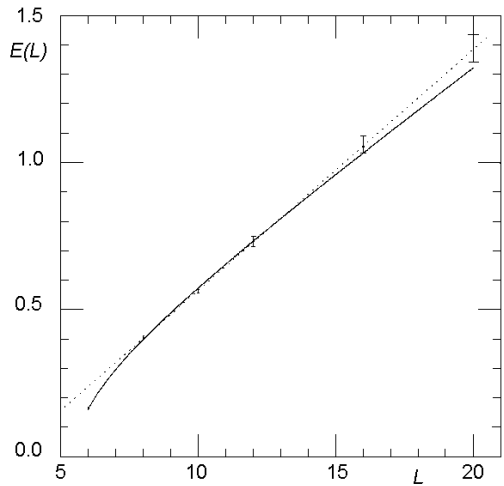

Figure 14: The torelon energy $E(L)$ versus $L$ for a periodic boundary of size $L$.

In conclusion the hadronic string model is a useful guide to the nature of the extended color flux but evidence for the characterization of self-energy from a particular type of string is not yet completely compelling.

## 9. Conclusions

We have discussed the status of the results of numerical measurements of the string tension in lattice gauge theories, in particular for $SU(2)$ and $SU(3)$. This relevant physical quantity has a good scaling behavior[126,60]. In order to make that explicit we have to use an improved coupling constant. In fact the string tension remains constant, for example in $SU(3)$, in the $\beta$ region $5.70 - 6.80$ where the lattice spacing changes by a factor 6. Although the quality of the agreement varies from one frame to the other, and a 20 percent discrepancy between different methods[69,60,63] remains, the approach to the expected continuum limit seems confirmed.

Present and future efforts are and will be directed to the study of larger lattices, at larger values of the coupling constant. The other main direction points to a better study of full QCD.

In fact the confinement in pure gauge QCD is visualized as the formation of a color electric flux-tube between two isolated static quarks. The strength of the linear contribution is connected to the string tension $\sigma$. This picture represented by the phenomenological spin-independent potential as given by the eq.5.3 changes when including dynamical quarks. At short distances dynamical quarks renormalize the gauge coupling changing the value of $\alpha_{q\bar{q}}(r)$. At long distances the creation of a $q - \bar{q}$ pair from the vacuum becomes favorable, and the string breaks.

This review has been discussing only quenched lattice gauge theories, but we note that lattice calculations that include some sea quark effects begin to exist (see for example ref. [182,180,181] and references therein); moreover a program to estimate these effects phenomenologically from quarkonium spectroscopy[177] has already given promising results for a full QCD determination of the strong coupling constant [179] including an even more realistic spectrum of sea quarks.

In this new context all the efforts to establish effective algorithms have to be replicated. One has to do a new tuning and to try and select the most efficient algorithms. The high $\beta$ limit will involve a fight against critical slowing down and small quark masses call for efficient methods to inverse large sparse matrices. For example in [183] we have discussed a very promising approach to the measure of properties related to confinement in fully coupled theories. We note that also the method determining directly the strong coupling from the interquark force[60] is applicable to full QCD.

### Acknowledgments

We are indebted to Giorgio Parisi for such a long lasting, enjoyable and productive collaboration. We thank Andreas Kronfeld for a correspondence containing useful advice.